\documentclass[10pt,journal,twoside]{IEEEtran}
\IEEEoverridecommandlockouts

\ifCLASSOPTIONcompsoc
\usepackage[nocompress]{cite}
\else
\usepackage{balance}
\usepackage{cite}
\fi
\ifCLASSINFOpdf
\usepackage[pdftex]{graphicx}
\graphicspath{{../Figures/}}
\DeclareGraphicsExtensions{.pdf,.jpeg,.png}
\else
\usepackage{graphicx}

\fi

\usepackage{amsthm}
\usepackage{amsmath,amssymb,amsfonts}
\usepackage{algorithmic}
\usepackage[ruled,vlined,linesnumbered]{algorithm2e}
\usepackage{smartdiagram}
\usepackage{geometry}
\usepackage{epsfig}
\usepackage{color}
\usepackage{lettrine}
\usepackage{float}
\usepackage{lipsum}
\usepackage{bm}
\usepackage{subcaption}
\usepackage{mathtools}
\usepackage{bbm}
\usepackage{etoolbox}
\usepackage{suffix}
\usepackage[T1]{fontenc}
\usepackage[colorlinks]{hyperref}
\usepackage[nameinlink,noabbrev]{cleveref}
\usepackage{textcomp}
\usepackage{xcolor}
\usepackage{multicol}
\usepackage{mdwmath}
\usepackage{mdwtab}
\usepackage{stackengine}
\usepackage{cancel}
\allowdisplaybreaks[3]

\usepackage{dsfont}

\newcommand{\x}{\mathbf{x}}

\newcommand{\ol}{\overline}

\definecolor{lime}{HTML}{A6CE39}
\DeclareRobustCommand{\orcidicon}{
\begin{tikzpicture}
\draw[lime, fill=lime] (0,0)
circle[radius=0.16]
node[white]{{\fontfamily{qag}\selectfont \tiny \.{I}D}};
\end{tikzpicture}
\hspace{-2mm}
}
\foreach \x in {A, ..., Z}{%
\expandafter\xdef\csname orcid\x\endcsname{\noexpand\href{https://orcid.org/\csname orcidauthor\x\endcsname}{\noexpand\orcidicon}}
}

\def\BibTeX{{\rm B\kern-.05em{\sc i\kern-.025em b}\kern-.08em
    T\kern-.1667em\lower.7ex\hbox{E}\kern-.125emX}}
\begin{document}

\title{Wormhole Detection Based on Z-Score And Neighbor Table Comparison}

\author{\IEEEauthorblockN{Zezhi Zeng\hspace{-1.5mm}\orcidB{}}
\vspace{-20pt}

\IEEEcompsocitemizethanks{
\IEEEcompsocthanksitem{Zezhi Zeng is
with the School of Cyberspace Security, Hainan University, Haikou, Hainan 570228, China
(e-mail: \{zzz918\}@hainanu.edu.cn).}
}
}

\maketitle
\begin{abstract}

Wormhole attacks can cause serious disruptions to the network topology in disaster rescue opportunity networks. By establishing false Wormhole(WH) links, malicious nodes can mislead legitimate paths in the network, further causing serious consequences such as traffic analysis attacks (i.e., by eavesdropping and monitoring exchanged traffic), denial of service (DoS) or selective packet loss attacks. This paper uses rescue equipment (vehicle-mounted base stations, rescue control centers, etc.) as an effective third-party auditor (TPA), and combines the commonly used Z-Score (Standard Score) data processing method to propose a new detection method based on pure mathematical statistics for detecting wormhole attacks. Finally, we perform a large number of simulations to evaluate the proposed method. Since our proposed strategy does not require auxiliary equipment such as GPS positioning and timers, as a pure data statistical analysis method, it is obviously more economically valuable, feasible, and practical than other strategies in disaster relief.
\end{abstract}

\begin{IEEEkeywords}
Opportunistic Network; Delay-tolerant Network; Disaster Relief; Wormhole
\end{IEEEkeywords}
\section{Introduction}
In this paper, we propose a new statistical detection method based on the commonly used Z-Score data processing method to efficiently detect wormhole attacks. Our results show that the proposed method can effectively detect wormhole attacks in opportunistic networks in disaster relief. Since our proposed strategy does not require auxiliary equipment such as GPS positioning and timers, as a pure data statistical analysis method, it is obviously more economically valuable, feasible and practical in disaster relief than other strategies.

By establishing false WH links, malicious nodes can mislead legitimate paths in the network, which in turn leads to a series of serious consequences, including traffic analysis attacks (such as eavesdropping and monitoring exchanged traffic), denial of service (DoS) attacks, or selective packet loss attacks. In disaster relief scenarios, communication resources are extremely scarce, and even slight tampering or misleading of the routing topology may cause serious consequences, not to mention the subsequent diversified network attacks. These consequences are unbearable for the victims waiting for assistance. 

Wormhole attacks are divided into exposed mode (internal attack, the attacker node identity is exposed, that is, the wormhole node is part of the network) and hidden mode (external attack, the attacker node identity is hidden)~\cite{Aslam2023}. In the exposed mode, the attacker nodes connected in pairs route packets using long-distance low-latency hidden links (through in-band or out-of-band channels)~\cite{Ala2024} and obtain high packet delivery rates, thereby deceiving legitimate nodes. When the packets arrive at the wormhole node, the attacker may launch traffic analysis, packet discard and/or packet tampering attacks. This chapter studies and analyzes the detection strategy of exposed mode wormhole attacks in opportunistic networks in disaster rescue scenarios to meet actual rescue situations.
Since the low latency of wormhole nodes is attractive to traffic (i.e., high packet delivery rate), we use the commonly used Z-Score data processing method to remove outliers and thus lock suspected wormhole nodes. Meanwhile, in a wormhole attack, a low-latency hidden channel (i.e., a wormhole tunnel) connects two strategically located attacker nodes, which are usually far away from each other, to capture more packets from neighboring nodes. This makes the nodes at both ends of the wormhole link look like neighbor nodes, as if they are actually adjacent to each other.

Taking rescue equipment (such as vehicle-mounted base stations, rescue control centers, etc.) as an effective third-party auditor (TPA), by comparing the distribution of neighbor nodes in the routing tables of both parties of the stripped suspected wormhole nodes, the nodes with low similarity are finally confirmed as wormhole nodes. The simulation experiment uses ONE Simulator to detect four major routing protocols in opportunistic networks (i.e., Prophet, Spray and Wait, Epidemic, and First Contact) to show the significant advantages of the proposed strategy.
The main contributions of this paper are as follows:
\begin{itemize}
    \item First, we constructed a wormhole model in the disaster rescue scenario of an opportunistic network and analyzed the harmfulness of wormholes to disaster rescue practice.
    \item Second, we combined the commonly used mathematical Z-Score and proposed a new statistical-based detection method for efficient detection of wormhole attacks.
    \item Finally, we evaluated our method through a large number of simulations. The results show that the proposed detection method has a high detection success rate (while reducing the false alarm rate) and significant application value in disaster rescue.
\end{itemize}

The rest of the paper is organized as follows. Section \ref{sec:Related} reviews related work. Section \ref{sec:model} introduces the system model. Section \ref{sec:z-Score} analyzes the system. Simulations are conducted in Section \ref{sec:Simulation} and Section \ref{sec:conclusion} concludes the paper.
\section{Related Work}\label{sec:Related}
This section provides an overview of the related works about wormhole in opportunistic network.

Wormholes (WH) aim to destroy the communication links between nodes. This type of attack is easy to launch and does not require encryption keys or knowledge of network protocols. It can seriously disrupt the network topology and is another major security attack in opportunistic networks. In terms of wormhole (WH) attacks, corresponding research results have also emerged, such as infrastructure-based defense mechanisms~\cite{Pham2014,Dhurandher2018}, geography-based detection methods~\cite{Ren2010}, trust-based defense methods~\cite{Liang2014}, strategies based on disabled topology~\cite{Jianing2018, Jyothirmai2015}, and some hybrid strategies.
An infrastructure-based WH defense mechanism was proposed~\cite{Pham2014}, which combines the statistical neighbor counting technology in DTN and the geography-based detection method proposed in the literature~\cite{Ren2010}. The infrastructure consists of two important regional distributions: first, 16 semi-distributed square regional networks within a range of 1 km (1 square kilometer); second, 16 nodes are deployed in the monitoring area. In each square, a node moves from one vertex to another for monitoring. The sensor node checks the status of adjacent nodes through neighbor counting technology (called neighbor number test) and performs all distance tests to detect changes in the shortest path length. This technology is divided into two phases (training phase and testing phase) to detect WH attacks, but it cannot locate the attacker node. In the training phase, the maximum average number of neighbor nodes is calculated and the threshold is set accordingly; in the testing phase, the difference between the maximum average number of nodes in a small area of the network area and the current number of nodes is calculated. If the number of neighbor nodes exceeds the maximum average threshold, it is determined to be a WH attack and the current neighbor node is notified. In 2018, an infrastructure-dependent method~\cite{Dhurandher2018} was proposed to detect malicious nodes (including WH, BH and MitM attacks) by mapping symmetric and asymmetric encryption technologies to trust-based routing algorithms for dense opportunistic networks. Reference~\cite{Liang2014} simulates the cooperation scenario and attack behavior of selfish nodes in DTN and proposes a trust-based epidemic routing protocol (TBER), which is a mechanism that can defend against WH attacks without relying on any infrastructure in DTN. In 2018, a WH attack detection mechanism based on maximum independent sets was proposed~\cite{Jianing2018}. It assumes that the transmission coverage of nodes follows the unit disk graph (UDG) communication model and sets a threshold for the transmission range of adjacent nodes. It also assumes that each node has at least two independent neighbors. The algorithm detects WH nodes by checking whether the node has two independent neighbors. If detected, it will be further processed to verify the WH node. Reference~\cite{Jyothirmai2015} proposed an algorithm for detecting WH attacks in DTN, which combines the Kruskal algorithm and the Euclidean distance and introduces a verifier in the technology based on the disabled topology.
Although these research results have promoted the development of this field to a certain extent, there are still many topics that need to be further studied in order to effectively improve network security protection capabilities. Most previous studies rely on auxiliary equipment (such as GPS positioning, timers, etc.) for design, and rarely consider the characteristics of opportunistic network nodes in resource-scarce disaster rescue scenarios. In order to meet the actual needs of disaster rescue scenarios, we will use the statistical method Z-Score (Standard Score) as a pure data statistical scheme to design a wormhole attack detection strategy for opportunistic networks in disaster rescue. Compared with other strategies, this solution has significant advantages in terms of economy, practicality and operability.
\section{System Modeling}\label{sec:model}

\begin{figure}
	\centering
	\includegraphics[width=0.95\columnwidth]{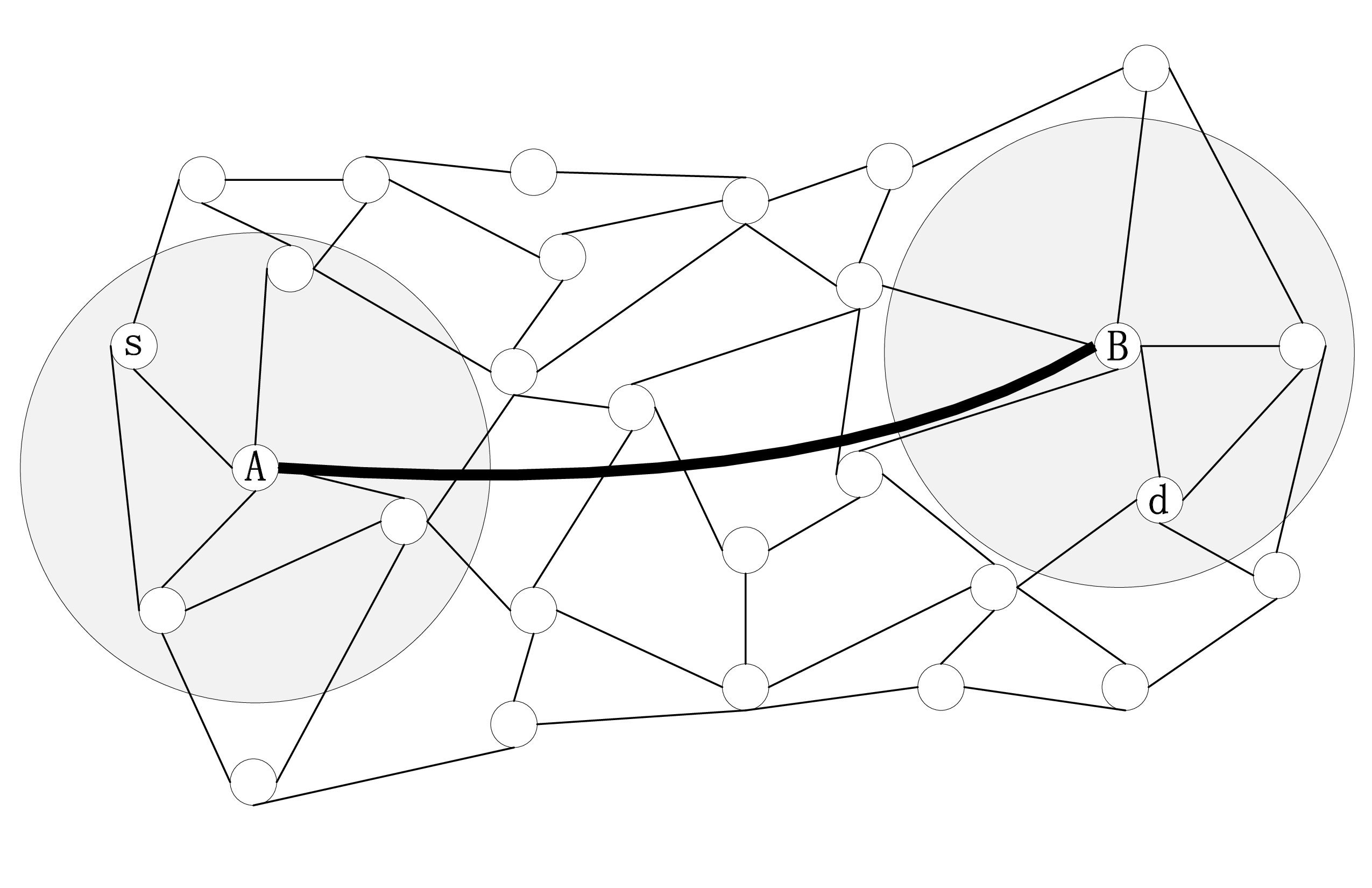}
	\caption{An Example of Wormhole in Network}
	\label{fig:wormhole}
    \vspace{-5pt}
\end{figure}

To facilitate the discussion of wormhole detection, the network model is first explained. Fig.~\ref{fig:wormhole} shows a typical wormhole example. In this scenario, nodes A and B are two malicious nodes deployed by the attacker, and they form the two endpoints of the wormhole link through a physical connection (such as a network cable). The data packets received by node A will be transmitted to node B through the wormhole link and replayed at node B, and vice versa.

Due to this mechanism, legitimate nodes near node A will mistakenly believe that nodes near node B are their neighbors, and vice versa. For example, the source node s can send a data packet directly to the target node d using a single-hop path through the wormhole link, without the need to transmit data in a multi-hop path under normal circumstances. This situation significantly interferes with the normal routing behavior of the network and facilitates subsequent attacks.

The typical feature of a wormhole attack is that the attacker eavesdrops on data packets at one end of the wormhole, transmits them to the other end through the wormhole tunnel, and then replays these data packets. By properly arranging wormhole node pairs, the attacker can attract as much routing traffic as possible in the network. Once the wormhole node is placed in a key position, a large number of network routes will be directed to the wormhole link, which will significantly disrupt the normal communication of the network.
\begin{figure*}[ht]
	\centering
	\includegraphics[width=1.3\columnwidth]{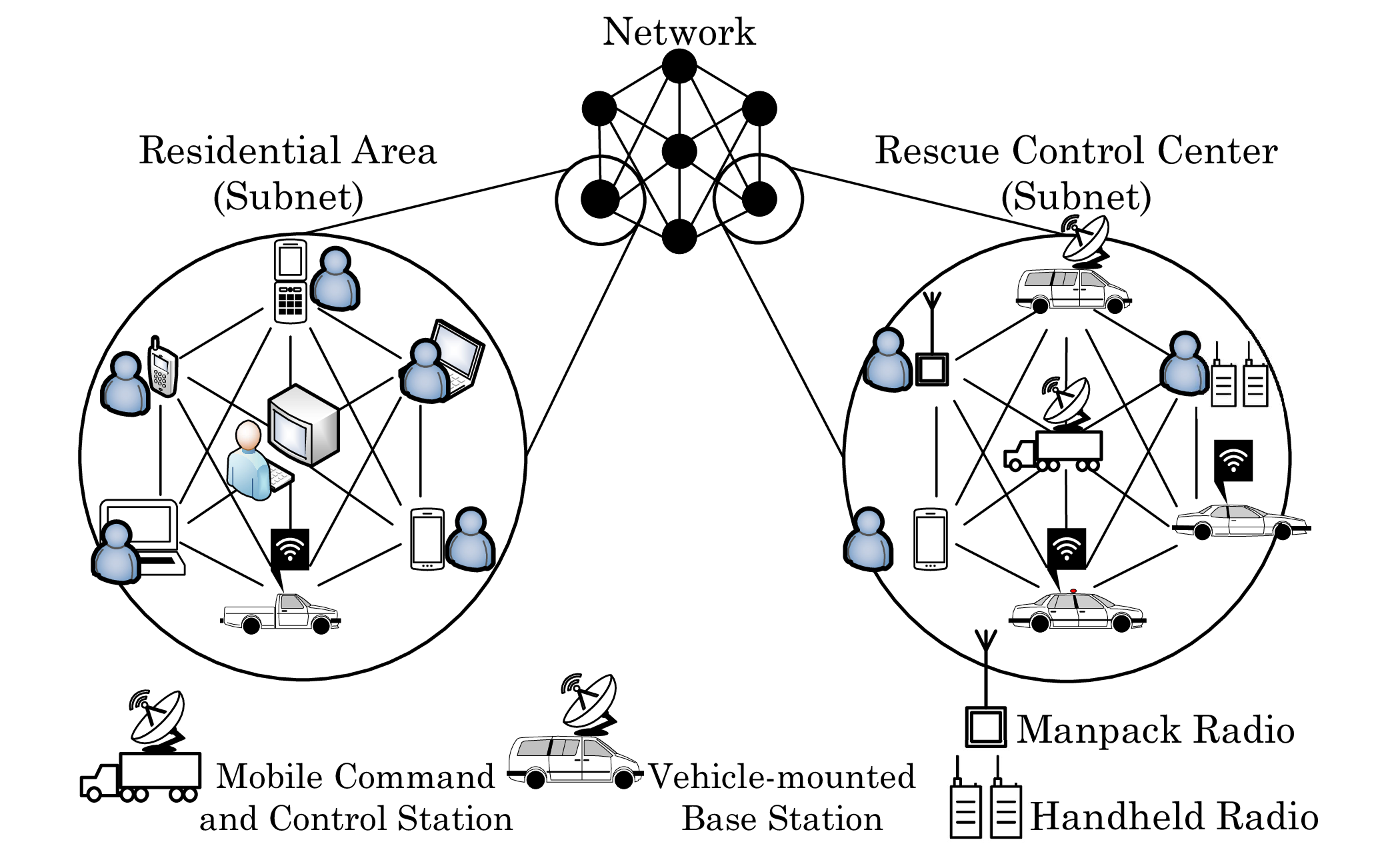}
	\caption{The Disaster Rescue Model for Network}
	\label{fig:subnet}
\vspace{-10pt}
\end{figure*}

\begin{figure}[ht]
	\centering
	\includegraphics[width=0.9\columnwidth]{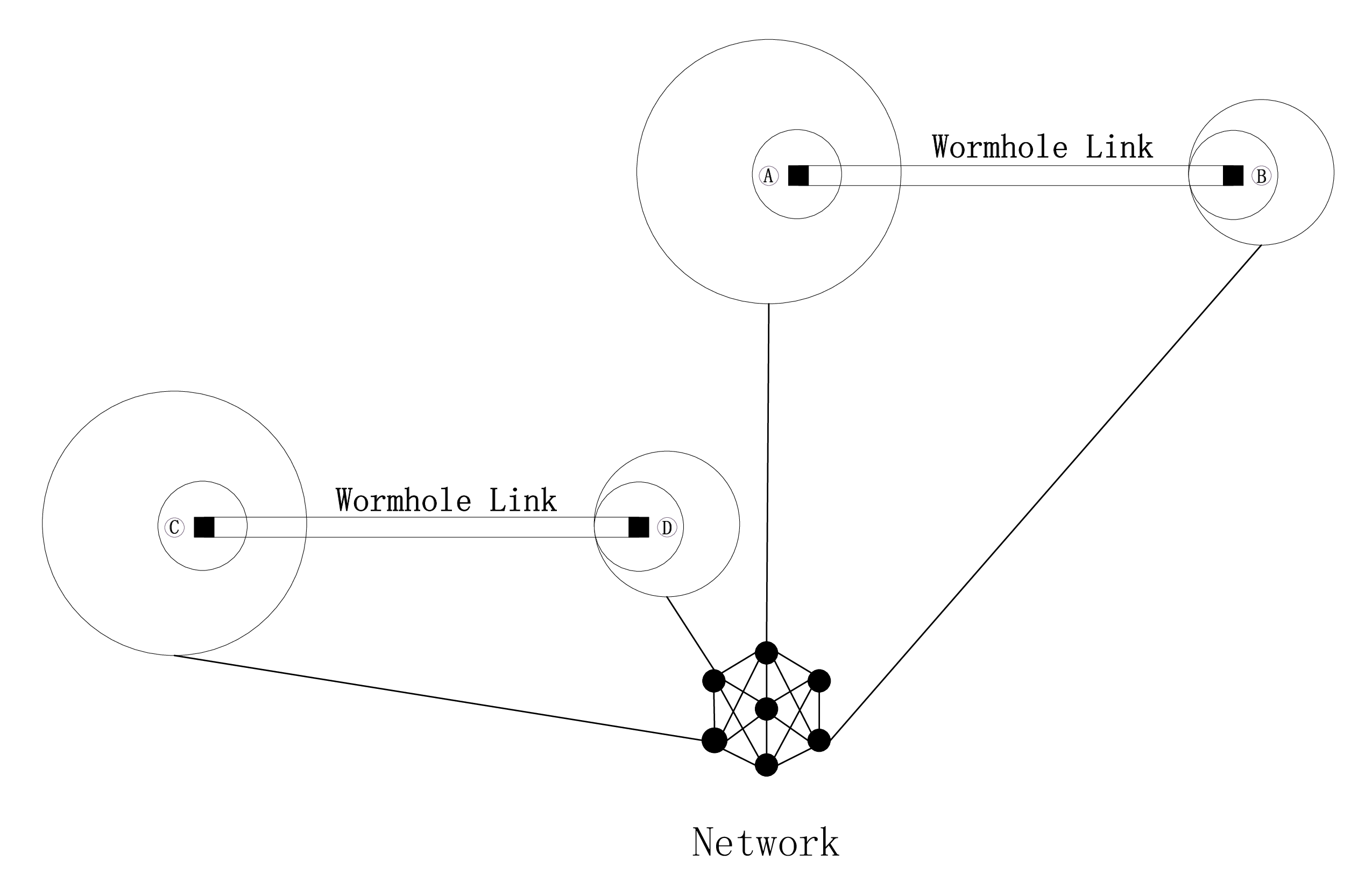}
	\caption{The Model of Wormhole Network}
	\label{fig:subnet_wormhole}
\vspace{-10pt}
\end{figure}

As shown in Fig.~\ref{fig:subnet}, in the disaster rescue scenario, in order to maximize the rescue efficiency and for the victims to survive, they will also gather at the rescue center where resources are concentrated. In fact, due to the clustering effect, the entire network is divided into multiple community-like areas, in which one or more rescue centers (rescue control centers, residential areas, temporary rescue stations) are reasonably deployed in the core area, and their huge gravitational effect will directly affect and guide the movement of nodes.
Moreover, wormhole node pairs will be placed in different areas (communities) far away from each other to achieve the purpose of attracting more traffic, which is also a typical feature of wormhole nodes. Wormhole nodes are also different from normal nodes in terms of transmission range, power and computing power (stronger capabilities).
Based on the above actual situation, the final wormhole network model is shown in Fig.~\ref{fig:subnet_wormhole}. It can be seen that the two pairs of nodes A and B, and C and D are two pairs of wormhole nodes. The node pairs are located in areas far apart, so as to achieve the purpose of obtaining more traffic. At the same time, they have better capabilities than ordinary nodes.

\section {Z-Score (Standard Score)}\label{sec:z-Score}
\subsection{Definition of Z-Score}
Z-Score~\cite{Yaro2024,Santhanakrishnan2022,Silva2024} is a commonly used statistical method for marking outliers. Through it, data of different orders of magnitude can be converted into a uniformly measured z-score value for comparison. The z-score represents the standard deviation between an element and the mean. Given an observation $x$, its z-score can be calculated by the following formula~\cite{Kreyszig2010}:
\begin{equation}
\begin{aligned}
    z= (x-\mu)/\sigma
    \label{z_score}
\end{aligned}
\end{equation}
Where $\mu$ and $\sigma$ represent the overall mean and standard deviation of $x$, respectively. The Z-Score truly reflects the relative standard distance from the mean of the score. If we convert each score into a z-score, each z-score will represent the distance or deviation from the mean of a specific score in the form of a standard deviation. The specific situation is as follows.

The Z-score is used to measure the position of a score relative to the mean. It indicates how many standard deviations the score differs from the mean. If a score is higher than the mean, its Z-score is positive; if it is lower than the mean, its Z-score is negative. Therefore, the Z-score can reflect the relative position of a value in the overall distribution.

Through the Z-score, we can intuitively see the degree of deviation of a data point from the mean. If all data points are converted to Z scores, each Z score will measure the distance of the point from the mean in units of standard deviation. In normal distribution data, the Z score transformation allows us to calculate the cumulative probability between the mean and the score by consulting the standard normal distribution table, thereby determining the percentile rank of the score in the data set. In addition, the sum of the squares of all Z scores in the data set is equal to the number of data points, and the standard deviation and variance of the Z score are always equal to 1.

\subsection{Improvement of Z-Score}
In order to improve the applicability of Z-Score on different data sets, the following improvements can be made:
\begin{itemize}
    \item Modified Z-Score: Use median and median absolute deviation (MAD) instead of mean and standard deviation to reduce the impact of extreme values:
    \begin{equation}
          \begin{aligned}
             Z^*= \frac{0.6745\cdot(x-median)}{MAD}
          \label{z-score}
          \end{aligned}
    \end{equation}
    Where: $Median$ is the median of the data (more robust than the mean and less affected by extreme values); $MAD$ (Median Absolute Deviation) is the median absolute deviation, calculated as: $MAD=\textbf{Median}(|x_i-\textbf{Median}|)$; $0.6745$ is an adjustment factor so that in the case of a normal distribution, $MZ$ and the standard Z-Score have similar threshold ranges.
    \item Local Z-Score: Calculate the mean and standard deviation for different data subsets separately, making the Z-Score more effective in non-uniform data distribution.
    \item Dynamic Z-Score: In streaming data scenarios, a sliding window is used to update the mean and standard deviation so that the Z-Score can adapt to real-time data changes.
\end{itemize}

\subsection{Selection of Z-Score}
Since the low latency of wormhole nodes has great appeal to traffic (i.e., high packet delivery rate), although the node data does not strictly follow the normal distribution, the standard Z-Score analysis method can still effectively remove the group value, and then lock the suspected wormhole node, laying the foundation for the final confirmation.

The z-score of the wormhole node is large. Even if the data is long-tailed, the extreme values will still deviate significantly from the mean, because the z-score of these points is far beyond the normal range and is an outlier. At the same time, our data volume is large enough, even if there is a certain deviation, the mean and standard deviation can still reflect the overall trend. Therefore, this paper uses the standard Z-Score statistical method to detect abnormal traffic of wormhole nodes.
\subsection{Comparison of neighbor tables}
In the network, each pair of nodes can establish a link, assuming that their distance is less than or equal to the transmission range $r$. For any node $m$, the set of neighbors of $m$ is represented by $N(m)$. For example, if node $B$ can receive a packet from node $A$ in one hop, then node $B$ is a neighbor of node $A$ and satisfies $B\in(A)$. The principle is to use the geometric relationship of node locations (transmission ranges) to check the neighbor topology under the communication range constraints of the two related nodes.

\begin{figure}[ht]
	\centering
	\includegraphics[width=1.1\columnwidth]{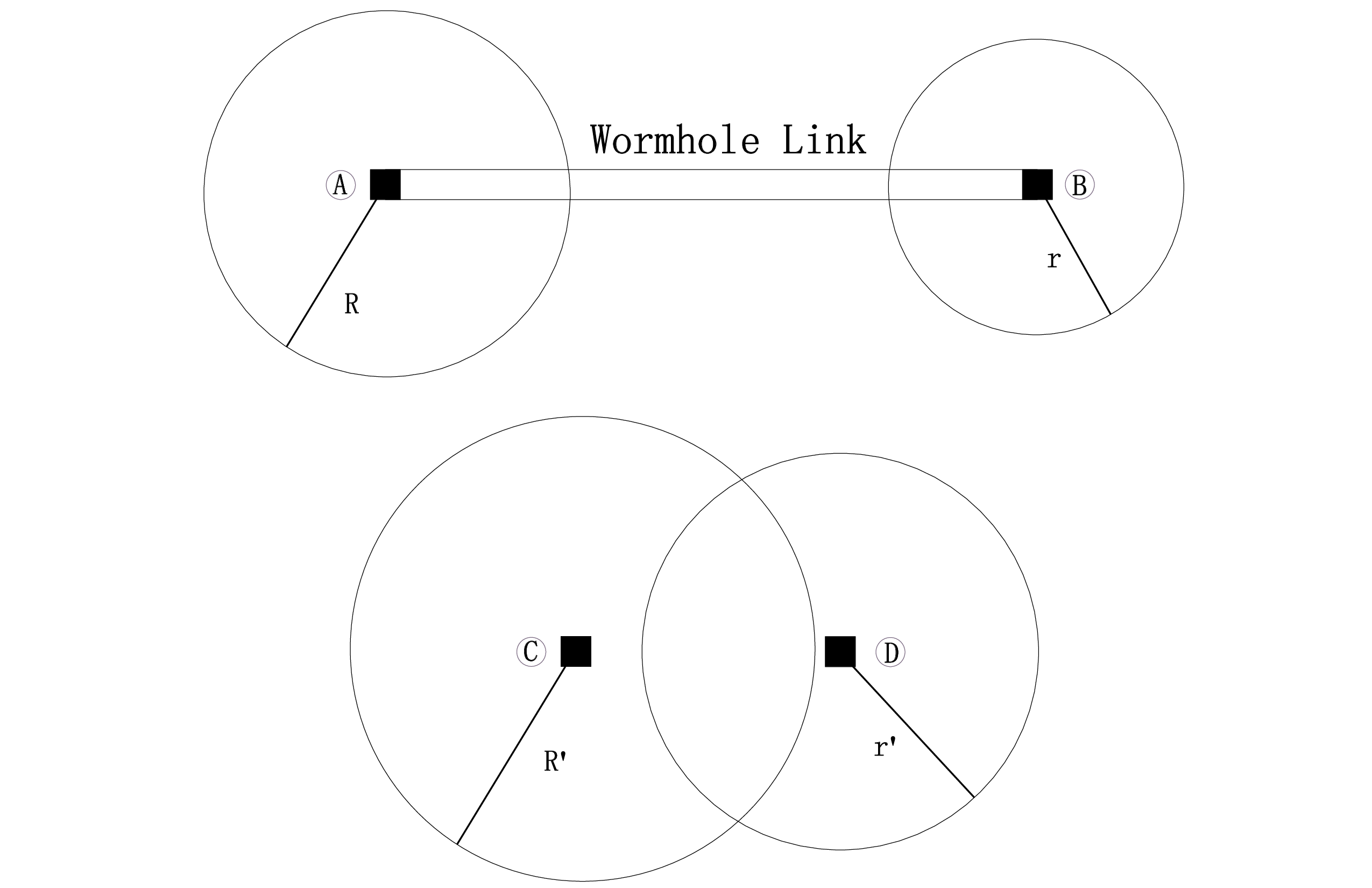}
	\caption{The Location Relationship of Different Type of Nodes}
	\label{fig:subnet_wormhole_2}
\vspace{-10pt}
\end{figure}

As shown in Fig.~\ref{fig:subnet_wormhole_2}, based on the network model constructed in the previous chapter, we have obtained two types of location distribution of suspected wormhole nodes:

(1) Wormhole node pairs: there are nodes $X\in N(A)$, $X\not\in N(B)$, satisfying $N(A)\not\subset N(B)$;

(2) Normal node pairs: there are nodes $Y\in N(C), Y\in N(D)$, satisfying $N(C) \subseteq N(D)$.

That is to say, since the wormhole nodes are far apart from each other, the inclusion relationship of their neighbor node sets is very weak. This is a very reliable evidence to confirm whether it is a wormhole node~\cite{Ahmadi2022}.

The complete wormhole detection strategy process is as follows:

\subsubsection{Routing Table}
In terms of routing tables, they can be aggregated to a third-party trusted institution by flooding or the incentive strategy in the previous chapter, and the routing information of suspected wormhole nodes should be derived and restored from the routing information of other normal nodes to prevent wormhole nodes from providing false routing tables.
\subsubsection{Communication Frequency}
Count the communication frequency of nodes, divide the nodes into suspected wormhole nodes and normal nodes, and mark the suspected wormhole nodes for further detection.
\subsubsection{Neighbor Nodes}
Count the routing table information of suspicious wormhole node pairs to compare the similarity of neighboring nodes. Nodes with extremely low similarity are considered wormhole nodes.
\section{Performance Evaluation}\label{sec:Simulation}
This chapter uses the simulator ONE (Opportunistic Networking Environment)~\cite{Ari2009, Nakayima2024} to perform simulations, test the four main routing protocols in opportunistic networks (i.e., Prophet, Spray and Wait, Epidemic, and First Contact), analyze and verify the performance of the proposed strategy.
\begin{table}
\captionsetup{font=footnotesize}
\caption{Notations and Values}
\centering
\resizebox{\columnwidth}{!}{%
	\begin{tabular}{|c|p{0.5\linewidth}|} 
		\hline
		\textbf{Notations} & \textbf{Values} \\ [0.5ex] 
		\hline 
            The range of simulation area & $4500*3400m^2$ \\
        \hline
            The number of nodes & $58,64,70,76$ \\ 
        \hline
            The number of wormhole nodes & $10 (5 pairs)$ \\
        \hline
		The time of simulation & $12h$ \\
        \hline

		The range of radio transmission (normal node) & $10m$ \\
        \hline
        The speed of node movement (normal node) & $0.5m/s$\~{}$1.5m/s;2.7m/s$\~{}$13.9m/s$ \\
        \hline
        The size of buffer (normal node) & $5M$ \\
        \hline
        The speed of data transmission (normal node) & $250k$ \\
        \hline
        The range of radio transmission (wormhole node) & $>500m$ \\
        \hline
        The speed of node movement (wormhole node) & $7m$\~{}$10m$ \\
        \hline
        The size of buffer (wormhole node) & $50M$ \\
        \hline
        The speed of data transmission (wormhole node) & $10M$ \\
        \hline
        The size of packet & $(500kB, 1MB)$ \\
        \hline
        The interval of message generation & $(25s, 35s)$ \\
        \hline
\end{tabular}}
\vspace{-10pt}
\label{table:sysParams}
\end{table}
\subsection{Evaluation Environment}
This paper uses the built-in map (Helsinki) of ONE for simulation experiments. Within the given communication area, the mobility model is a random waypoint, the nodes are set to $58, 64, 70$, and $76$ respectively, the number of wormhole nodes is set to $5$ pairs, the total simulation time is $12h$, $10m$ is the communication transmission range of the normal node, $0.5m/s$\~{}$1.5m/s;2.7m/s$\~{}$13.9m/s$ is the moving speed of the normal node, the normal node buffer size is $5MB$, the normal node data transmission speed is set to $250k$, the signal coverage range of the wormhole node is $>500m$, $7m$\~{}$10m$ is the moving speed of the wormhole node, the wormhole node buffer size is $50MB$, the wormhole node data transmission speed is set to $10M$, the data packet size is $(500kB, 1MB)$, and the generation interval of each message is $25s$\~{}$35s$. The specific parameters are shown in Table~\ref{table:sysParams}.
\subsection{Network Performance Indicators}
In order to better analyze the strategy proposed in this paper. This section uses three important indicators: detection success rate, false alarm rate, and detection time to analyze the strategy.

(1) $\text{Detection Success Rate}=\frac{\text{Number of Detected Nodes}}{\text{Preset Number}}*100\%$, the success of detection is a very important indicator to measure whether the strategy is successful. The higher the detection success rate, the better the performance of the strategy;

(2) $\text{False {Alarm Rate}}= \frac{\text{Number of False Alarms}}{\text{Preset Number}}*100\%$, the false alarm rate is the proportion of nodes that were originally thought to be detected but are actually normal nodes to the preset wormhole nodes. The lower the false alarm rate, the stronger the robustness of the strategy;

(3) Detection time is the average time from the start of the attack to the declaration of the existence of the wormhole.

\subsection{Results in Evaluation}
\begin{figure}
    \centering
    \begin{subfigure}{0.25\textwidth}
    	\centering
    	\includegraphics[width=\linewidth]{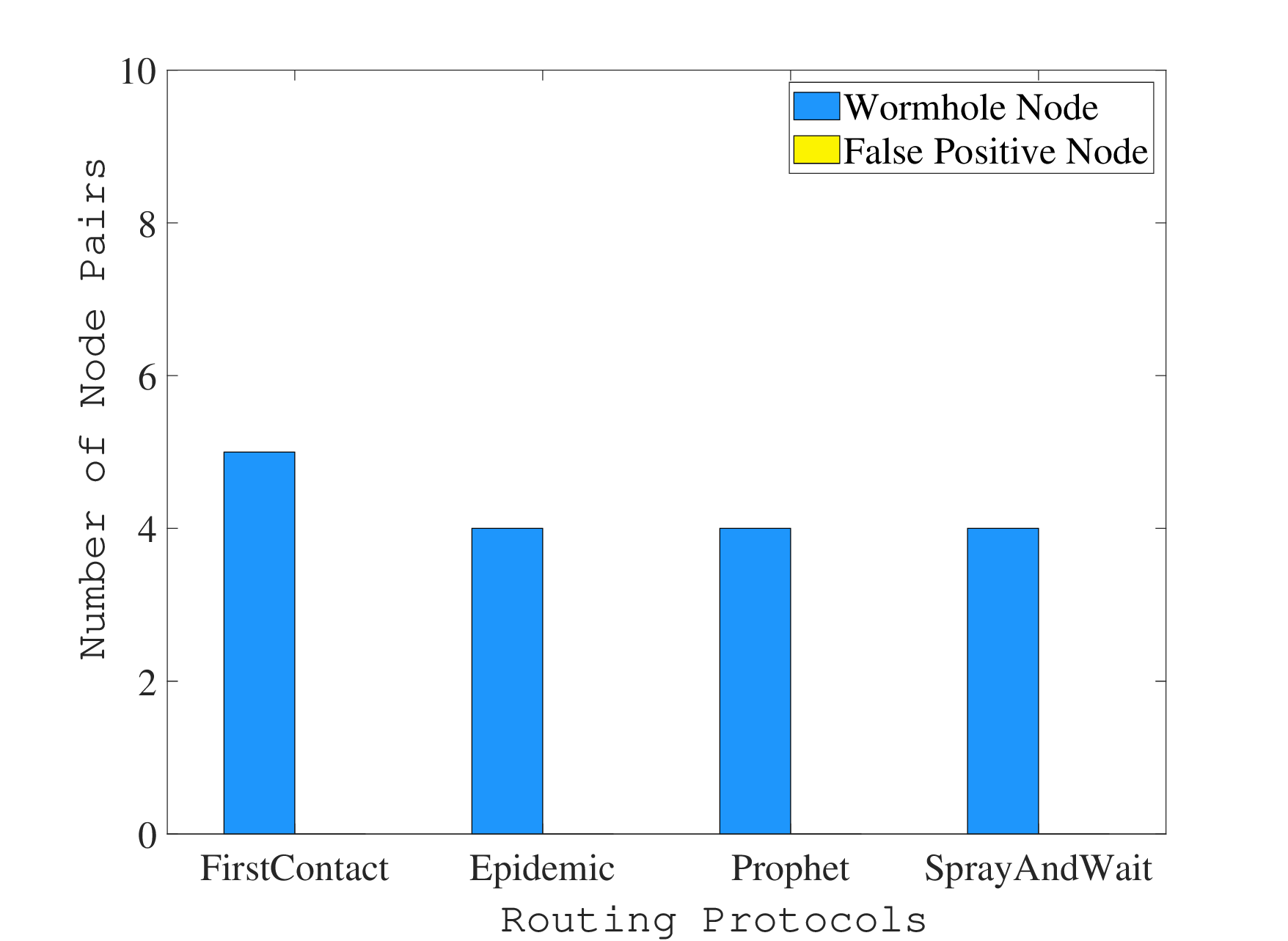}
    	\caption{}
    	\label{sim:fig5a}
    \end{subfigure}%
    \begin{subfigure}{0.25\textwidth}
    	\centering
    	\includegraphics[width=\linewidth]{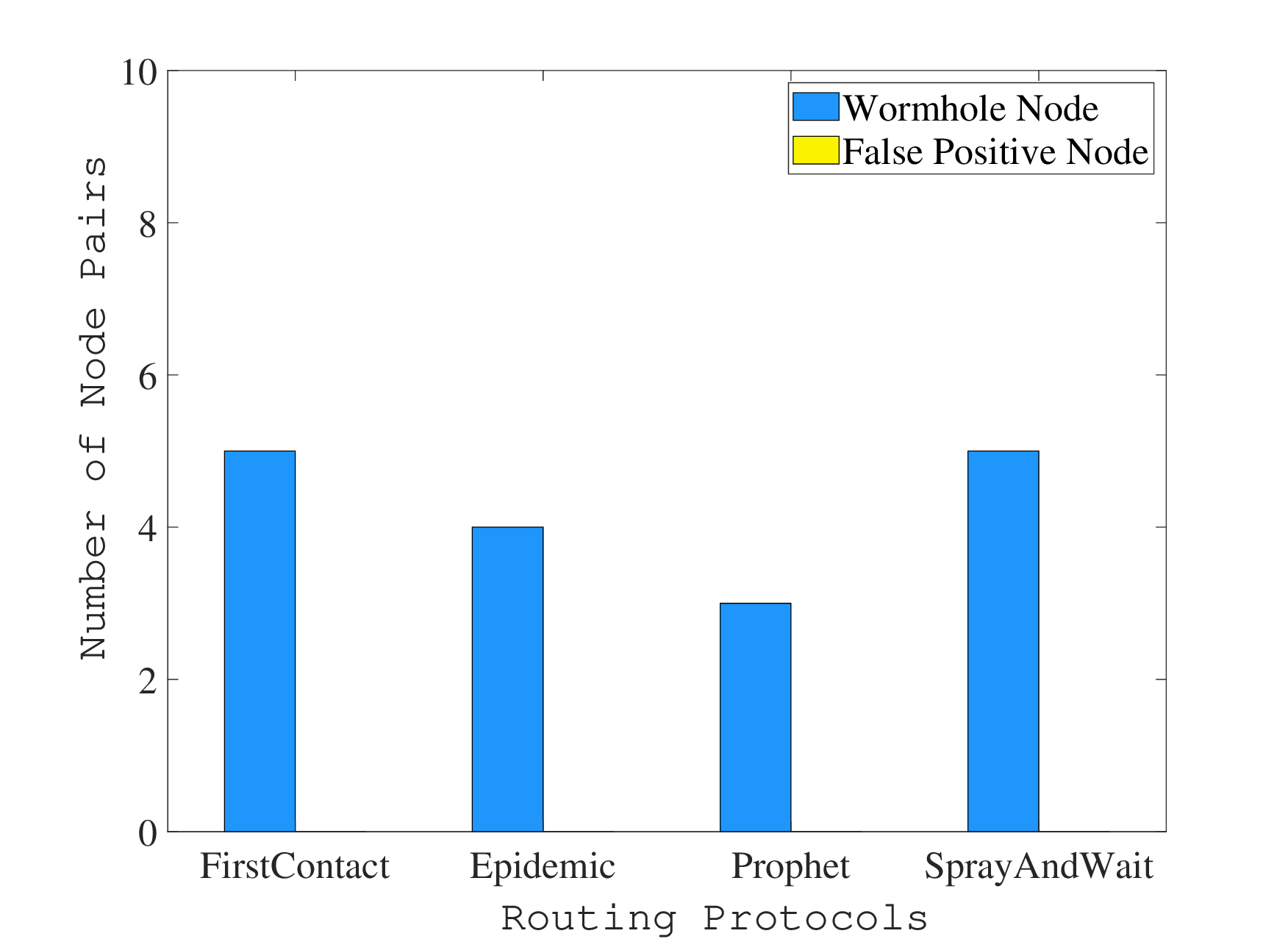}
    	\caption{}
    	\label{sim:fig5b}
    \end{subfigure}%
    \\
    \begin{subfigure}{0.25\textwidth}
    	\centering
    	\includegraphics[width=\linewidth]{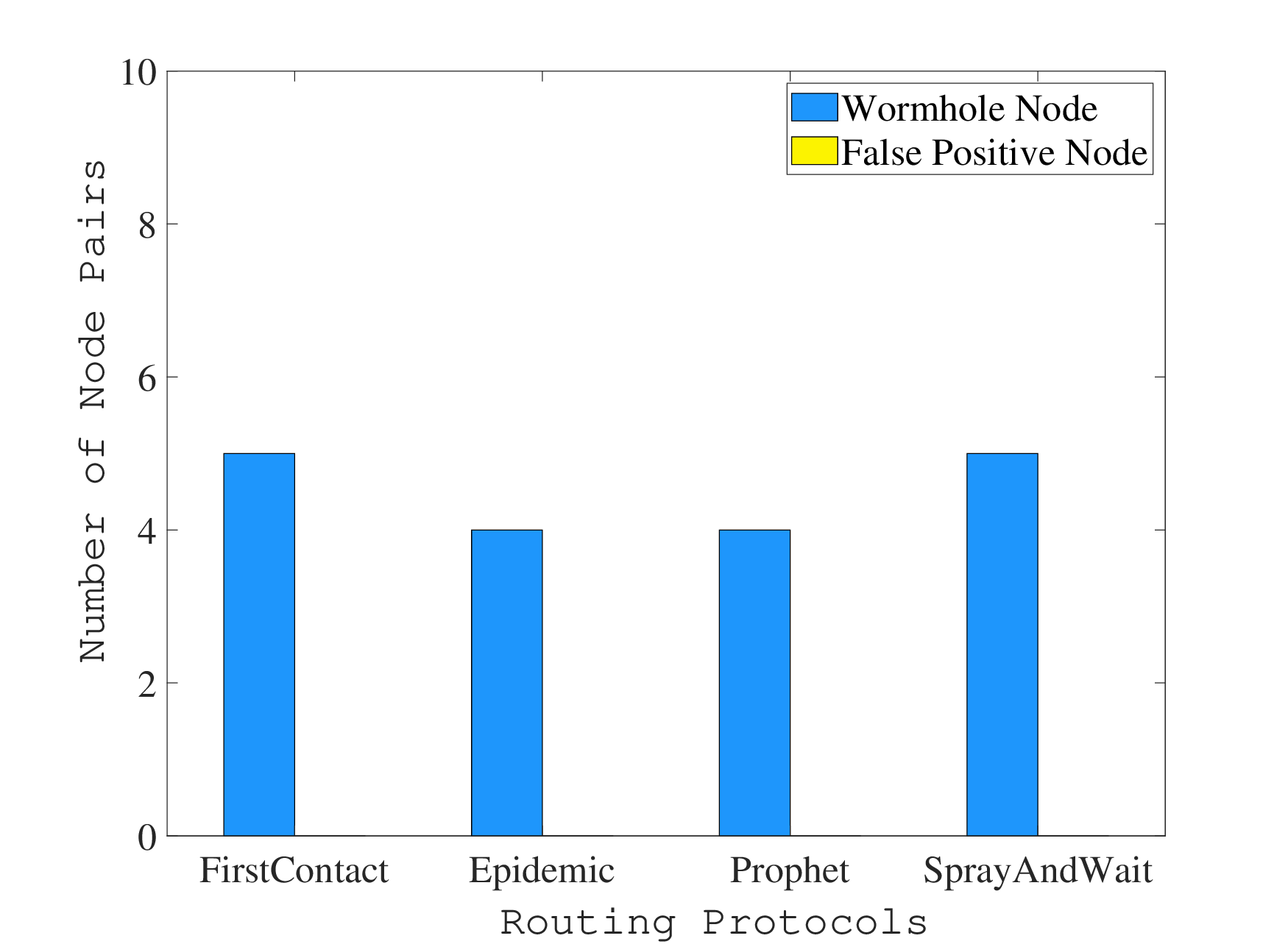}
    	\caption{}
    	\label{sim:fig5c}
    \end{subfigure}%
    \begin{subfigure}{0.25\textwidth}
    	\centering
    	\includegraphics[width=\linewidth]{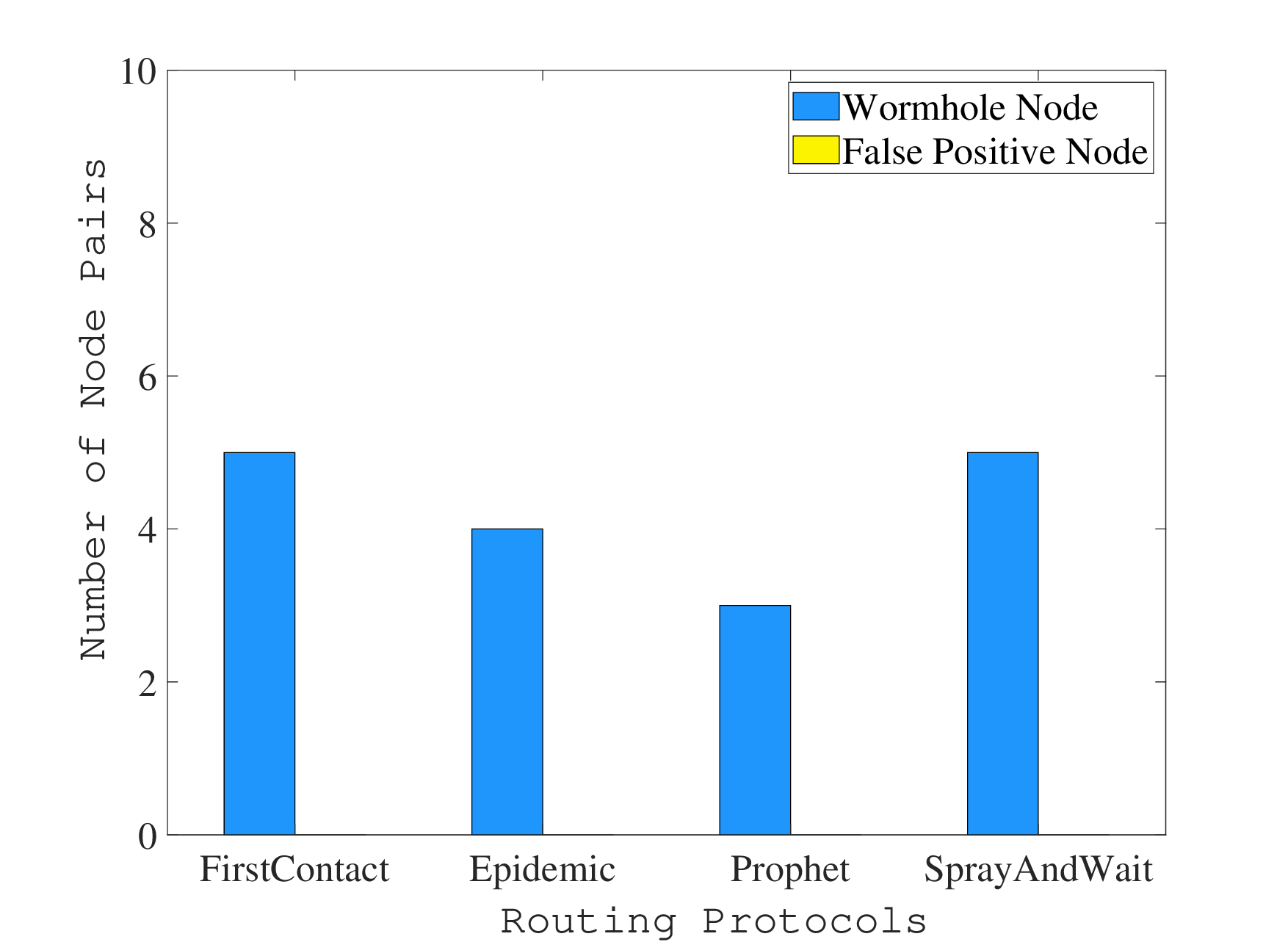}
    	\caption{}
    	\label{sim:fig5d}
    \end{subfigure}%

    \caption{Details Detected when Node Density (58/64/70/76) increases Successively}
    \label{sim:wormhole}
\vspace{-10pt}
\end{figure}

In this section, the performance of the wormhole detection strategy proposed in this section can be verified by analyzing the simulation results, and compared with other strategies~\cite{Ren2010, Pham2014, Jianing2018}. The simulation uses the average detection success rate and the average false alarm rate as analysis indicators. Table~\ref{table:details} shows some simulation data. Fig.~\ref{sim:wormhole} shows an intuitive bar chart of the corresponding data.

\begin{table}
\captionsetup{font=footnotesize}
\caption{Details of Detections (\checkmark Wormhole nodes $\times$ False positive nodes)}
\centering
\begin{subtable}{0.2\textwidth}
    	\centering
    	\resizebox{\columnwidth}{!}{%
	\begin{tabular}{|c|c|c|} 
		\hline
		\textbf{Router} & \textbf{\checkmark} & \textbf{$\times$} \\ [0.8ex] 
		\hline 
            FirstContact & $5$ & $0$\\
        \hline
            Epidemic & $4$ & $0$ \\ 
        \hline
             Prophet & $4$ & $0$ \\
        \hline
		SprayAndWait & $4$ & $0$ \\
        \hline

       \end{tabular}}
    \end{subtable}%
   \begin{subtable}{0.2\textwidth}
    	\centering
    	\resizebox{\columnwidth}{!}{%
	\begin{tabular}{|c|c|c|} 
		\hline
		\textbf{Router} & \textbf{\checkmark} & \textbf{$\times$} \\ [0.8ex] 
		\hline 
            FirstContact & $5$ & $0$\\
        \hline
            Epidemic & $4$ & $0$ \\ 
        \hline
             Prophet & $3$ & $0$ \\
        \hline
		SprayAndWait & $5$ & $0$ \\
        \hline

       \end{tabular}}
    \end{subtable}%
    \\
   \begin{subtable}{0.2\textwidth}
    	\centering
    	\resizebox{\columnwidth}{!}{%
	\begin{tabular}{|c|c|c|} 
		\hline
		\textbf{Router} & \textbf{\checkmark} & \textbf{$\times$} \\ [0.8ex] 
		\hline 
            FirstContact & $5$ & $0$\\
        \hline
            Epidemic & $4$ & $0$ \\ 
        \hline
             Prophet & $4$ & $0$ \\
        \hline
		SprayAndWait & $5$ & $0$ \\
        \hline

       \end{tabular}}
    \end{subtable}%
    \begin{subtable}{0.2\textwidth}
    	\centering
    	\resizebox{\columnwidth}{!}{%
	\begin{tabular}{|c|c|c|} 
		\hline
		\textbf{Router} & \textbf{\checkmark} & \textbf{$\times$} \\ [0.8ex] 
		\hline 
            FirstContact & $5$ & $0$\\
        \hline
            Epidemic & $4$ & $0$ \\ 
        \hline
             Prophet & $3$ & $0$ \\
        \hline
		SprayAndWait & $5$ & $0$ \\
        \hline

       \end{tabular}}
    \end{subtable}%

\vspace{-10pt}
\label{table:details}
\end{table}

\begin{figure}[ht]
	\centering
	\includegraphics[width=1.0\columnwidth]{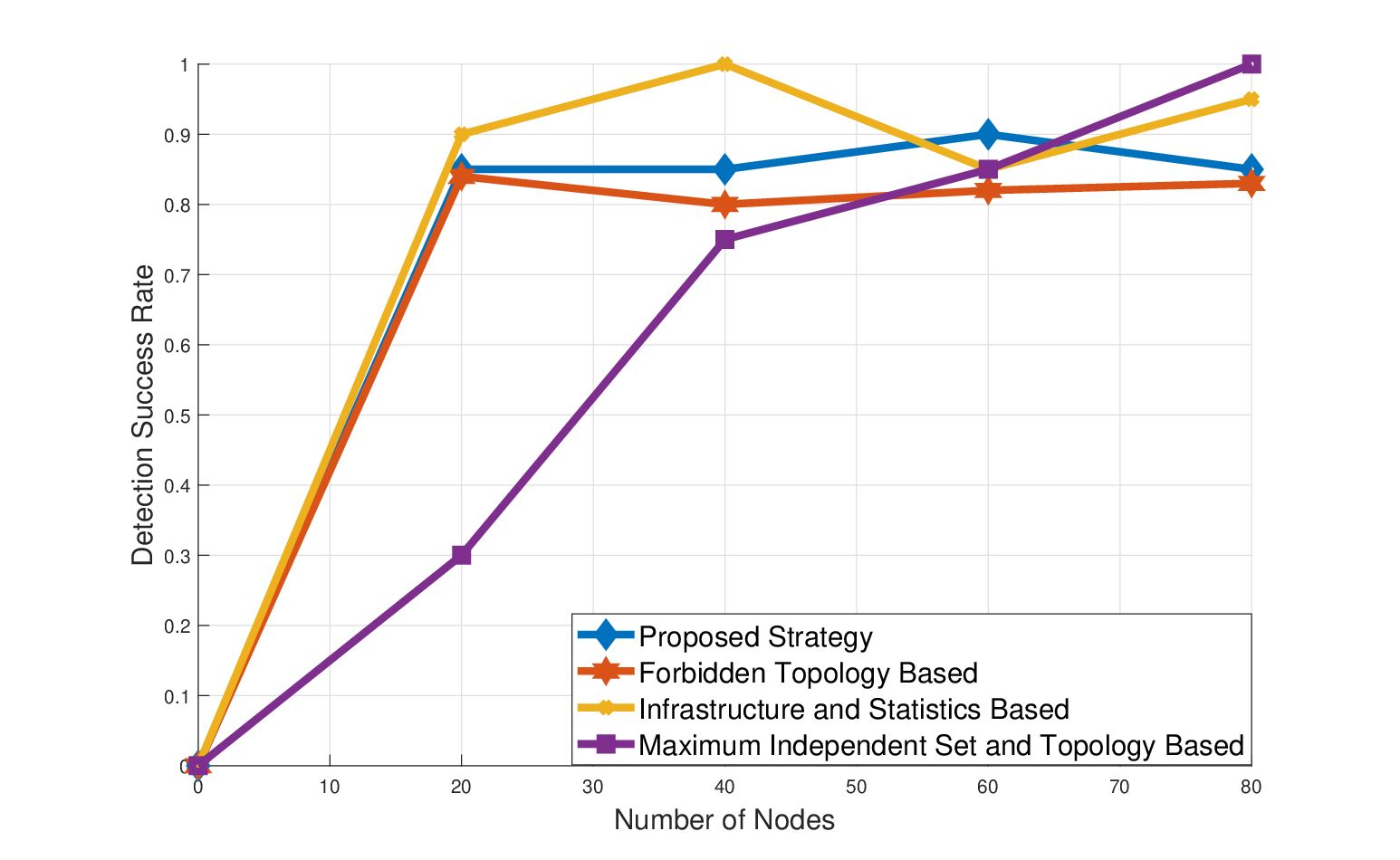}
	\caption{Detection Success Rate of Different Strategies}
	\label{fig:success}
\vspace{-10pt}
\end{figure}

As shown in Fig.~\ref{fig:success}, by comparing with the existing strategies, the proposed method is comparable to other methods in terms of detection accuracy. It can be seen that the proposed method has a relatively stable detection result, which also reflects that the wormhole attack has obvious characteristics in mathematical statistics. The wormhole has a relatively single attack mode, that is, frequently transmitting data packets to each other through a shared private hidden channel. This is what distinguishes the wormhole from other attack methods. The binding relationship between wormhole node pairs is also relatively fixed. However, in terms of detection speed and false alarm rate, the proposed method has obvious advantages.

\begin{figure}[ht]
	\centering
	\includegraphics[width=1.0\columnwidth]{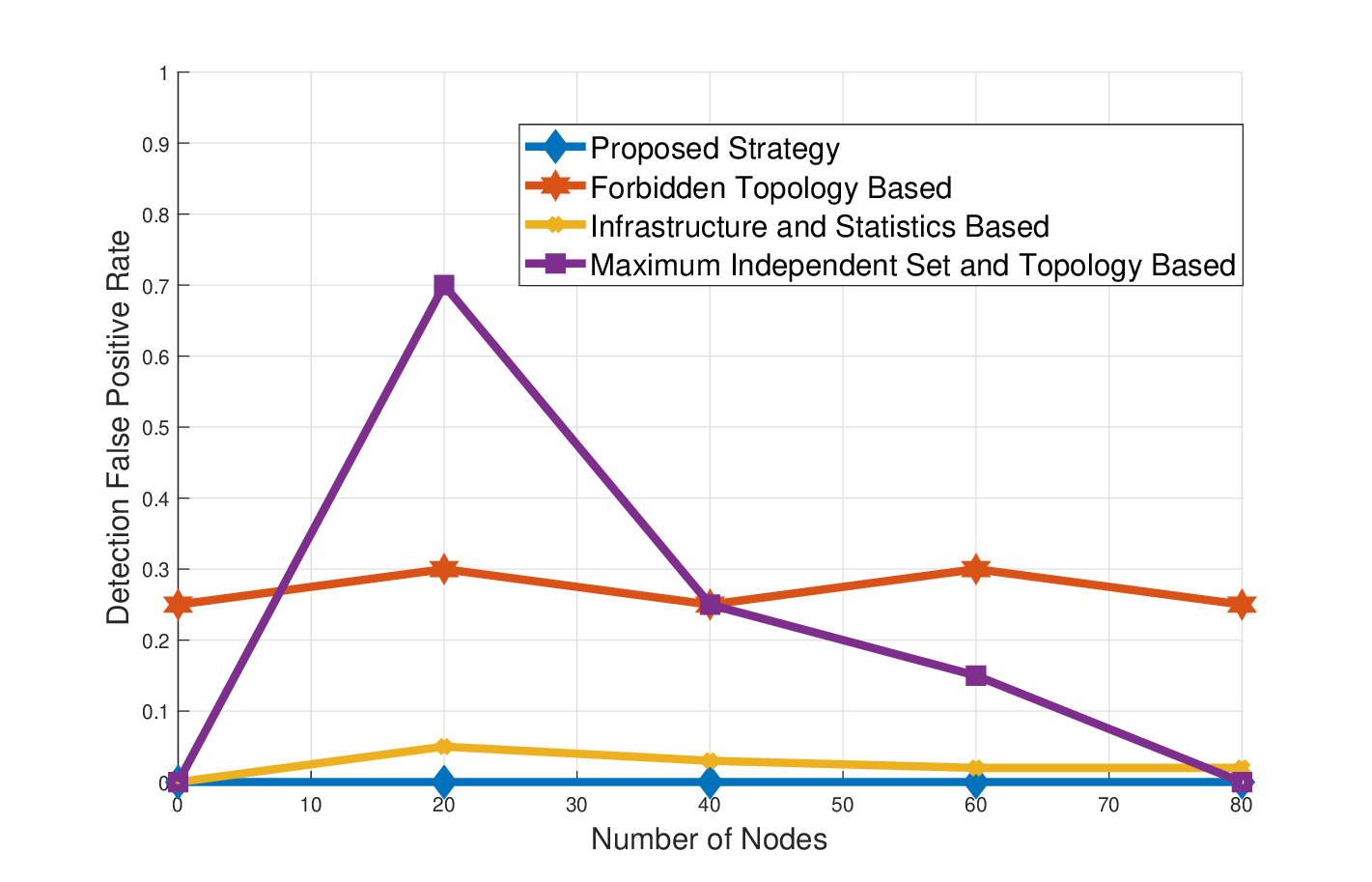}
	\caption{Detection Success Rate of Different Strategy}
	\label{fig:positive}
\vspace{-10pt}
\end{figure}

\begin{figure}[ht]
	\centering
	\includegraphics[width=1.0\columnwidth]{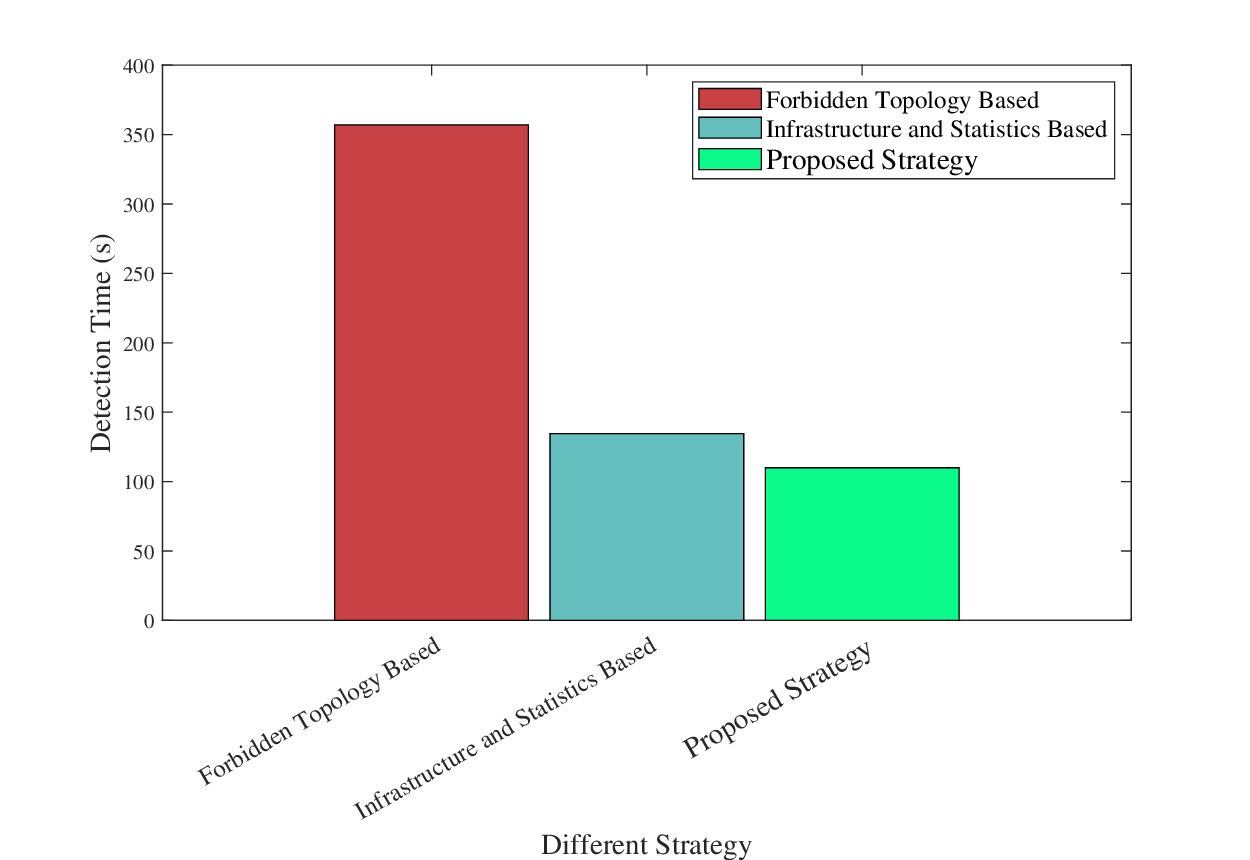}
	\caption{Detection Time of Different Strategy}
	\label{fig:detecttime}
\vspace{-10pt}
\end{figure}

As shown in Fig.~\ref{fig:positive} and Fig.~\ref{fig:detecttime}, it can be clearly seen that the false alarm rate is the biggest advantage of the method in this paper. No matter how the number of nodes changes, the proposed method has an extremely low false alarm rate. This is because in the opportunistic network, only wormhole nodes with hidden links can show huge traffic flow, that is, rapid filtering of data packets, while normal nodes do not have this ability to obtain traffic and are far less efficient than wormhole nodes. At the same time, the detection speed of the proposed method is also relatively ideal, which reflects the importance of the rescue equipment as a third-party auditing agency to quickly coordinate and configure resources. The behavior of the node is recorded at all times, and the traces of the crime cannot be erased. The system can provide feedback on the attack behavior in real time. It should be pointed out that the effect of wormhole positioning mainly depends on the detection performance. The higher the detection success rate, the lower the false alarm rate, the higher the positioning accuracy, and the lower the positioning false alarm rate. For example, the forbidden topology method in the literature~\cite{Ren2010} assumes the use of the unit disk graph (UDG) model and relies on the distance between nodes to judge the neighbor relationship. However, in practice, the rapid movement of nodes and the propagation loss or collision of the network MAC layer may lead to inaccurate maintenance of the neighbor list, thereby increasing the false alarm rate. In this chapter, the rescue equipment, as a third-party trusted agency, not only assumes the relay function but also has the audit function. This enables the rescue equipment to effectively and quickly maintain the neighbor list and start the detection program, thereby improving the response efficiency of the network.

As the transmission range and the number of nodes increase, the number of neighbors of each node will also increase. The forbidden topology method~\cite{Ren2010} requires three nodes to fall into the wormhole link to detect anomalies, which means that the more neighbors there are, the greater the possibility of forming a taboo structure, which can detect wormholes faster and improve the accuracy of detection. The infrastructure and statistical strategies in reference~\cite{Pham2014} have the problem of requiring additional infrastructure, and the functions of auxiliary infrastructure must be integrated into the existing mesh network deployment. However, the method in reference ~\cite{Pham2014} is based on the increasing trend of the number of neighbors in a small area, and can detect the proportional difference before and after the wormhole appears by observing the change of the number of neighbors over time, without relying on the absolute value of the number of neighbors in the area.

A problem with the method based on maximum independent sets and topology in reference~\cite{Jianing2018} is that the detection effect can only be fully exerted when the number of nodes reaches a certain level, that is, the detection success rate and accuracy gradually increase as the number of nodes increases. This has certain disadvantages when the node density is sparse, especially in remote areas of disaster relief where the number of nodes is obviously insufficient.

\begin{figure}[ht]
	\centering
	\includegraphics[width=1.0\columnwidth]{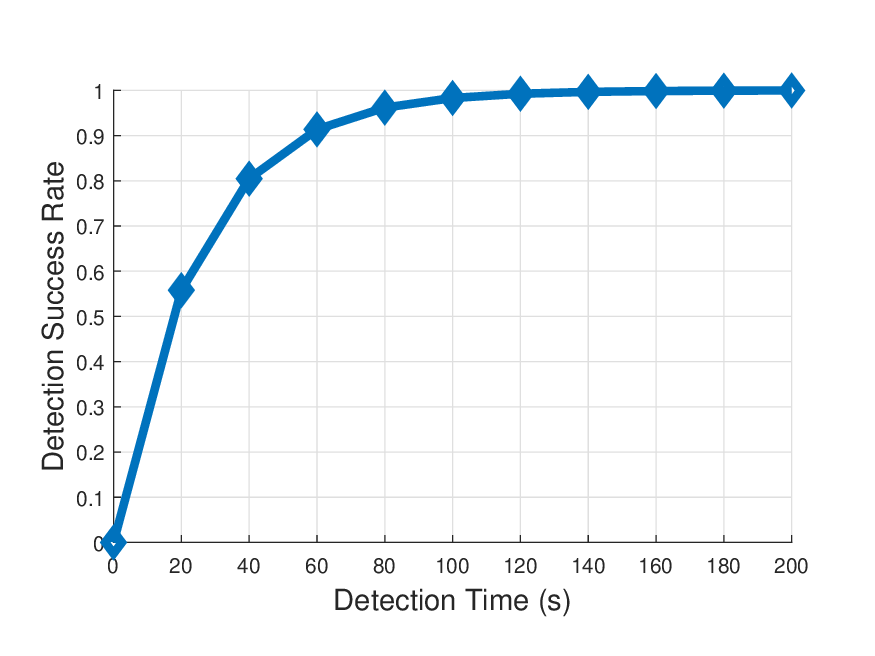}
	\caption{Detection Success Rate over Time}
	\label{fig:success rate}
\vspace{-10pt}
\end{figure}

It can be seen from the experimental results in Fig.~\ref{fig:success rate} that as the detection time increases, the longer the node stays in one location, the more neighbor counting evidence it can collect, so that the existence of the wormhole can be judged more accurately and confidently. It should be pointed out that in our solution, the rescue equipment, as a third-party trusted institution, has both the relay node function and the execution of the detection program, that is, the detection starts when the system is started, but it does not need to be monitored all the time. It can wait until the system becomes stable, that is, when the behavior pattern of the wormhole attack begins to emerge, the system is preheated, and then it is necessary to execute the detection program.

\section{Conclusions}\label{sec:conclusion}
Wormhole attacks are one of the most serious security attacks that can be encountered in opportunistic networks, aiming to destroy the communication links between nodes. Such attacks are easy to launch and do not require encryption keys or knowledge of network protocols. By establishing false WH links, malicious nodes can mislead legitimate paths in the network, further causing serious consequences such as traffic analysis attacks. Many protocols have been proposed to defend against wormhole attacks in wireless networks by using synchronized clocks, positioning devices, or directional antennas. In this paper, we introduce a new method for detecting wormhole attacks, which is a wormhole detection strategy based on the Z-Score statistics and the correlation of neighbor nodes. Unlike many existing techniques, it does not use any dedicated hardware, making it very useful in practical scenarios. However, most importantly, no matter how large the transmission range is, the algorithm can always prevent wormholes by checking the local neighborhood information to determine whether the network topology is true or false, and the efficiency of the algorithm will not be affected even if the topology changes dynamically, which shows the robustness of the algorithm.

\ifCLASSOPTIONcaptionsoff
\newpage
\fi

\bibliographystyle{IEEEtran}
\bibliography{myReferences}

\begin{thebibliography}{10}
\providecommand{\url}[1]{#1}
\csname url@samestyle\endcsname
\providecommand{\newblock}{\relax}
\providecommand{\bibinfo}[2]{#2}
\providecommand{\BIBentrySTDinterwordspacing}{\spaceskip=0pt\relax}
\providecommand{\BIBentryALTinterwordstretchfactor}{4}
\providecommand{\BIBentryALTinterwordspacing}{\spaceskip=\fontdimen2\font plus
\BIBentryALTinterwordstretchfactor\fontdimen3\font minus \fontdimen4\font\relax}
\providecommand{\BIBforeignlanguage}[2]{{%
\expandafter\ifx\csname l@#1\endcsname\relax
\typeout{** WARNING: IEEEtran.bst: No hyphenation pattern has been}%
\typeout{** loaded for the language `#1'. Using the pattern for}%
\typeout{** the default language instead.}%
\else
\language=\csname l@#1\endcsname
\fi
#2}}
\providecommand{\BIBdecl}{\relax}
\BIBdecl

\bibitem{Aslam2023}
S.~Aslam, A.~Altaweel, and I.~Kamel, ``Exposed-mode of wormhole attack in opportunistic mobile networks: Impact study and analysis,'' in \emph{Proceedings of the 2023 European Interdisciplinary Cybersecurity Conference}, ser. EICC '23.\hskip 1em plus 0.5em minus 0.4em\relax New York, NY, USA: Association for Computing Machinery, 2023, p. 19–25.

\bibitem{Ala2024}
A.~Altaweel, S.~Aslam, and I.~Kamel, ``Security attacks in opportunistic mobile networks: A systematic literature review,'' \emph{Journal of Network and Computer Applications}, vol. 221, p. 103782, 2024.

\bibitem{Pham2014}
T.~N.~D. Pham and C.~K. Yeo, ``Statistical wormhole detection and localization in delay tolerant networks,'' in \emph{2014 IEEE 11th Consumer Communications and Networking Conference (CCNC)}, 2014, pp. 380--385.

\bibitem{Dhurandher2018}
S.~K. Dhurandher, A.~Kumar, and M.~S. Obaidat, ``Cryptography-based misbehavior detection and trust control mechanism for opportunistic network systems,'' \emph{IEEE Systems Journal}, vol.~12, no.~4, pp. 3191--3202, 2018.

\bibitem{Ren2010}
Y.~Ren, M.~C. Chuah, J.~Yang, and Y.~Chen, ``Detecting wormhole attacks in delay-tolerant networks [security and privacy in emerging wireless networks],'' \emph{IEEE Wireless Communications}, vol.~17, no.~5, pp. 36--42, 2010.

\bibitem{Liang2014}
X.~Liang, J.~Qin, M.~Wang, D.~Wang, and J.~Wan, ``An effective and secure epidemic routing for disruption-tolerant networks,'' in \emph{2014 Sixth International Conference on Intelligent Human-Machine Systems and Cybernetics}, vol.~2, 2014, pp. 329--333.

\bibitem{Jianing2018}
J.~Li, Q.~Wang, and Z.~Gao, ``An improved detecting mechanism against wormhole attacks in delay tolerant networks,'' in \emph{2018 10th International Conference on Wireless Communications and Signal Processing (WCSP)}, 2018, pp. 1--6.

\bibitem{Jyothirmai2015}
P.~Jyothirmai and J.~S. Raj, ``Secure interoperable architecture construction for overlay networks,'' in \emph{2015 International Conference on Innovations in Information, Embedded and Communication Systems (ICIIECS)}, 2015, pp. 1--6.

\bibitem{Yaro2024}
A.~S. Yaro, F.~Maly, P.~Prazak, and K.~Malý, ``Outlier detection performance of a modified z-score method in time-series rss observation with hybrid scale estimators,'' \emph{IEEE Access}, vol.~12, pp. 12\,785--12\,796, 2024.

\bibitem{Santhanakrishnan2022}
K.~Santhanakrishnan and V.~Senthooran, ``A parallel distributed cluster computing model for z-score computation in respect of sri lankan university admissions,'' in \emph{2022 2nd International Conference on Advanced Research in Computing (ICARC)}, 2022, pp. 344--348.

\bibitem{Silva2024}
L.~H.~T. Bandória, W.~N. Silva, and M.~C. de~Almeida, ``Modeling uncertainties in electricity consumption using the z-score approach,'' in \emph{2024 Workshop on Communication Networks and Power Systems (WCNPS)}, 2024, pp. 1--7.

\bibitem{Kreyszig2010}
E.~Kreyszig, H.~Kreyszig, and E.~J. Norminton, \emph{Advanced Engineering Mathematics}, 10th~ed.\hskip 1em plus 0.5em minus 0.4em\relax Hoboken, NJ, USA: Wiley, 2010, p. 1040.

\bibitem{Ahmadi2022}
S.~Al-Ahmadi, W.~Aliady, and A.~AlRashedy, ``An efficient wormhole attack detection method in wireless sensor networks,'' in \emph{2022 26th International Conference on Circuits, Systems, Communications and Computers (CSCC)}, 2022, pp. 240--249.

\bibitem{Ari2009}
A.~Ker\"{a}nen, J.~Ott, and T.~K\"{a}rkk\"{a}inen, ``The one simulator for dtn protocol evaluation,'' in \emph{Proceedings of the 2nd International Conference on Simulation Tools and Techniques}, ser. Simutools '09.\hskip 1em plus 0.5em minus 0.4em\relax Brussels, BEL: ICST (Institute for Computer Sciences, Social-Informatics and Telecommunications Engineering), 2009:1-10.

\bibitem{Nakayima2024}
O.~Nakayima, M.~I. Soliman, K.~Ueda, and S.~A.~E. Mohamed, ``Combining software-defined and delay-tolerant networking concepts with deep reinforcement learning technology to enhance vehicular networks,'' \emph{IEEE Open Journal of Vehicular Technology}, vol.~5, pp. 721--736, 2024.

\end{thebibliography}

\end{document}